\documentstyle[12pt]{article}

\catcode`\@=11
\long\def\@makefntext#1{
\protect\noindent \hbox to 3.2pt {\hskip-.9pt
$^{{\ninerm\@thefnmark}}$\hfil}#1\hfill}		

\def\@makefnmark{\hbox to 0pt{$^{\@thefnmark}$\hss}}  

\def\ps@myheadings{\let\@mkboth\@gobbletwo
\def\@oddhead{\hbox{}
\rightmark\hfil\ninerm\thepage}
\def\@oddfoot{}\def\@evenhead{\ninerm\thepage\hfil
\leftmark\hbox{}}\def\@evenfoot{}
\def\sectionmark##1{}\def\subsectionmark##1{}}

\renewcommand{\thefootnote}{\fnsymbol{footnote}}

\newcounter{sectionc}\newcounter{subsectionc}\newcounter{subsubsectionc}
\renewcommand{\section}[1] {\vspace*{0.6cm}\addtocounter{sectionc}{1}
\setcounter{subsectionc}{0}\setcounter{subsubsectionc}{0}\noindent
	{\normalsize\bf\thesectionc. #1}\par\vspace*{0.4cm}}
\renewcommand{\subsection}[1] {\vspace*{0.6cm}\addtocounter{subsectionc}{1}
	\setcounter{subsubsectionc}{0}\noindent
	{\normalsize\it\thesectionc.\thesubsectionc. #1}\par\vspace*{0.4cm}}
\renewcommand{\subsubsection}[1]
{\vspace*{0.6cm}\addtocounter{subsubsectionc}{1}
	\noindent {\normalsize\rm\thesectionc.\thesubsectionc.\thesubsubsectionc.
	#1}\par\vspace*{0.4cm}}

\newcounter{appendixc}
\newcounter{subappendixc}[appendixc]
\newcounter{subsubappendixc}[subappendixc]

\renewcommand{\appendix}[1] {\vspace*{0.6cm}
        \refstepcounter{appendixc}
        \setcounter{figure}{0}
        \setcounter{table}{0}
        \setcounter{equation}{0}
        \renewcommand{\thefigure}{\Alph{appendixc}.\arabic{figure}}
        \renewcommand{\thetable}{\Alph{appendixc}.\arabic{table}}
        \renewcommand{\theappendixc}{\Alph{appendixc}}
        \renewcommand{\theequation}{\Alph{appendixc}.\arabic{equation}}
        \noindent{\bf Appendix \theappendixc #1}\par\vspace*{0.4cm}}

\def\abstracts#1{{

\centering{\begin{minipage}{12.2truecm}\footnotesize\baselineskip=12pt\noindent
	\centerline{\footnotesize ABSTRACT}\vspace*{0.3cm}
	\parindent=0pt #1
	\end{minipage}}\par}}

\renewenvironment{thebibliography}[1]
	{\begin{list}{\arabic{enumi}.}
	{\usecounter{enumi}\setlength{\parsep}{0pt}
\setlength{\leftmargin 1.25cm}{\rightmargin 0pt}
	 \setlength{\itemsep}{0pt} \settowidth
	{\labelwidth}{#1.}\sloppy}}{\end{list}}

\newcounter{itemlistc}
\newcounter{romanlistc}
\newcounter{alphlistc}
\newcounter{arabiclistc}

\newcommand{\fcaption}[1]{
        \refstepcounter{figure}
        \setbox\@tempboxa = \hbox{\footnotesize Fig.~\thefigure. #1}
        \ifdim \wd\@tempboxa > 6in
           {\begin{center}
        \parbox{6in}{\footnotesize\baselineskip=12pt Fig.~\thefigure. #1}
            \end{center}}
        \else
             {\begin{center}
             {\footnotesize Fig.~\thefigure. #1}
              \end{center}}
        \fi}

\newcommand{\tcaption}[1]{
        \refstepcounter{table}
        \setbox\@tempboxa = \hbox{\footnotesize Table~\thetable. #1}
        \ifdim \wd\@tempboxa > 6in
           {\begin{center}
        \parbox{6in}{\footnotesize\baselineskip=12pt Table~\thetable. #1}
            \end{center}}
        \else
             {\begin{center}
             {\footnotesize Table~\thetable. #1}
              \end{center}}
        \fi}

\def\@citex[#1]#2{\if@filesw\immediate\write\@auxout
	{\string\citation{#2}}\fi
\def\@citea{}\@cite{\@for\@citeb:=#2\do
	{\@citea\def\@citea{,}\@ifundefined
	{b@\@citeb}{{\bf ?}\@warning
	{Citation `\@citeb' on page \thepage \space undefined}}
	{\csname b@\@citeb\endcsname}}}{#1}}

\newif\if@cghi
\def\cite{\@cghitrue\@ifnextchar [{\@tempswatrue
	\@citex}{\@tempswafalse\@citex[]}}
\def\citelow{\@cghifalse\@ifnextchar [{\@tempswatrue
	\@citex}{\@tempswafalse\@citex[]}}
\def\@cite#1#2{{$\null^{#1}$\if@tempswa\typeout
	{IJCGA warning: optional citation argument
	ignored: `#2'} \fi}}

 1
 1
 1

\font\ninerm=cmr9

\input{psfig}

\begin{document}

\def\alt{\stackrel{<}{\sim}}
\def\agt{\stackrel{>}{\sim}}
\def\eslt{E\llap/_T}
\def\etmiss{E\llap/_T}
\def\tg{\tilde g}
\def\tst{\tilde t}
\def\tq{\tilde q}
\def\tf{\tilde f}
\def\tl{\tilde \ell}
\def\tell{\tilde \ell}
\def\tnu{\tilde \nu}
\def\te{\tilde e}
\def\tmu{\tilde \mu}
\def\ttau{\tilde \tau}
\def\tz{\widetilde Z}
\def\tw{\widetilde W}

\newcommand{\st}{\scriptstyle}
\newcommand{\sst}{\scriptscriptstyle}
\newcommand{\mco}{\multicolumn}
\newcommand{\epp}{\epsilon^{\prime}}
\newcommand{\vep}{\varepsilon}
\newcommand{\ra}{\rightarrow}
\newcommand{\ppg}{\pi^+\pi^-\gamma}
\newcommand{\vp}{{\bf p}}
\newcommand{\ko}{K^0}
\newcommand{\kb}{\bar{K^0}}
\newcommand{\al}{\alpha}
\newcommand{\ab}{\bar{\alpha}}
\def\be{\begin{equation}}
\def\ee{\end{equation}}
\def\bea{\begin{eqnarray}}
\def\eea{\end{eqnarray}}
\def\CPbar{\hbox{{\rm CP}\hskip-1.80em{/}}}

\hfill{FSU-HEP-950311}

\centerline{\normalsize\bf THE SEARCH FOR SUPERSYMMETRY}
\baselineskip=22pt

\centerline{\footnotesize HOWARD BAER}
\baselineskip=13pt
\centerline{\footnotesize\it Dep't of Physics, Florida State University}
\baselineskip=12pt
\centerline{\footnotesize\it Tallahassee, FL 32306, USA}
\centerline{\footnotesize E-mail: baer@hep.fsu.edu}
\vspace*{0.3cm}

\vspace*{0.9cm}
\abstracts{The minimal supergravity model (SUGRA),
with gauge coupling unification and
radiative electroweak symmetry breaking, is a well-motivated paradigm for
physics Beyond the Standard Model. In this talk, I review the
capabilities of various present and future collider experiments to make
definitive tests of the SUGRA model, including LEP and LEP II, the
Tevatron and its upgrades, and finally the CERN LHC project.}

\normalsize\baselineskip=15pt
\setcounter{footnote}{0}
\renewcommand{\thefootnote}{\alph{footnote}}
\section{Introduction}

When Dirac first incorporated Lorentz symmetry into quantum mechanics, he
found that every known particle had to have a ``partner'' particle, the
anti-particle. When supersymmetry, which is a natural generalization of the
Lorentz group, is incorporated into the framework of quantum field theory,
once again, all known particles are found to have partners: the so-called
sparticles.
Although supersymmetry was ``discovered'' in a mathematical sense in the late
1960's and 1970's, a supersymmetric version of the Standard Model (SM) did not
emerge until the early 1980's$^1$. The resulting theory, often called the
Minimal Supersymmetric Standard Model (MSSM)$^2$, is looked upon as
a likely low energy effective theory of some more fundamental superstring or
supergravity GUT model. Recent precision measurements of gauge couplings at
LEP,
when extrapolated to ultra-high energy scales via renormalization group
equations, acheive unification at a scale
$M_X\sim 2\times 10^{16}$ GeV in the MSSM, but not in the SM$^3$. This
fact has led to intense scrutinization of supersymmetric models with
unification.

The minimal supergravity model is ``minimal'' in the sense that only the
minimal number of new particles and interactions are introduced to be
consistent with phenomenology. In particular, this implies that $R$-parity
is conserved.
It is ``super''-symmetric,
in that the Lagrangian possesses supersymmetry, and hence predicts
super-partners.
The supersymmetry is broken by including soft-supersymmetry breaking terms,
which parametrize our ignorance of the exact mechanism of supersymmetry
breaking. ``Gravity'' enters by assuming a simple structure
for the soft-supersymmetry breaking terms at or beyond the unification scale.
For instance, if supersymmetry
is broken in a hidden sector of the model at very high energy scales, the
breaking can be communicated to the observable sector by gravitational
interactions, which are univeral in strength.
This leads to a common mass $m_0$ for all scalar soft-breaking
terms, a common gaugino mass $m_{1/2}$ for the three soft-breaking gaugino
masses, and common trilinear couplings $A_0$ and a bilinear term $B_0$.
Starting from the unification scale, the various couplings and masses are
evolved via renormalization group equations, which yield the low energy
(weak-scale) running sparticle masses and couplings. A consequence of this
mechanism is that electroweak symmetry is broken when one of the Higgs mass
squared terms is driven negative. Ultimately, the whole low energy
superparticle spectrum can be calculated in terms of the parameter set
\begin{equation}
m_0,\ m_{1/2},\ A_0,\ \tan\beta ,
\end{equation}
along with the sign of the superpotential Higgs mass $\mu$ and the top
mass $m_t$.

\section{Connection to experiment}

The above framework is simple, well-motivated, compelling, consistent
with data, and testable, since the soft supersymmetry breaking terms are
expected to be generally of order the weak scale: $\sim 100-1000$ GeV.
The key to connecting the SUGRA framework with experiment lies in the
event generator program. Recently, the MSSM, and in addition, the more
constrained SUGRA framework, has been incorporated into the event
generator program ISAJET$^4$,
so that super-particle production and decay can be
simulated for $e^+e^-$, $p\bar p$ or $pp$ colliders.

To simulate sparticle production and decay at a hadron collider, the following
steps are taken using ISAJET 7.13:
\begin{itemize}
\item input the parameter set ($m_0, m_{1/2},A_0,\tan\beta ,sign(\mu ), m_t$),
(or a less constrained MSSM set),
\item all sparticle and Higgs masses and couplings are computed,
\item all sparticle, top and Higgs decay modes and branching fractions
are calculated,
\item all lowest order $2\rightarrow 2$ sparticle production processes
are calculated (if desired, subsets of the reactions can be selected),
\item the hard scattering is convoluted with parton distribution functions,
\item initial and final state QCD radiation is calculated with the parton
shower
model,
\item particles and sparticles decay through their various cascades,
\item quarks and gluons are hadronized, and heavy hadrons are decayed,
\item the underlying soft scattering of beam remnants is modelled,
\item the resulting event and event history is generated for interface
with detector simulations, or for direct analysis.
\end{itemize}
For $e^+e^-$ collisions, the steps are similar, except that initial state
QCD radiation, PDF convolution and beam jet evolution do not enter.

\section{Sparticles at LEP and LEP II}

The four LEP experiments have placed relatively model independent
bounds of the masses of
various sparticles and Higgs bosons$^5$. To be specific$^6$,
\begin{eqnarray*}
 m_{\tw_1}&>&45\ {\rm GeV}, \\
 m_{\tz_1}&>&18\ {\rm GeV},\ ({\rm GUT\ relation\ assumed}),  \\
 m_{\tell}&>&45\ {\rm GeV},\ (\tell =\te ,\tmu , \ttau ),  \\
 m_{\tq}&>&45\ {\rm GeV},  \\
 m_{\tnu}&>&41.8\ {\rm GeV},\ ({\rm three\ degenerate\ flavors}),  \\
 m_{H_{\ell}}&\agt &60\ {\rm GeV}.
\end{eqnarray*}
The above sparticle mass limits are mainly
limited by the beam energy. Hence, considerable improvement is soon expected
from the energy upgrade to LEP II, running at $\sqrt{s}=150-190$ GeV,
at integrated luminosity of $300-500\ pb^{-1}$ per year.
\begin{figure}[tbh]
\centerline{\psfig{file=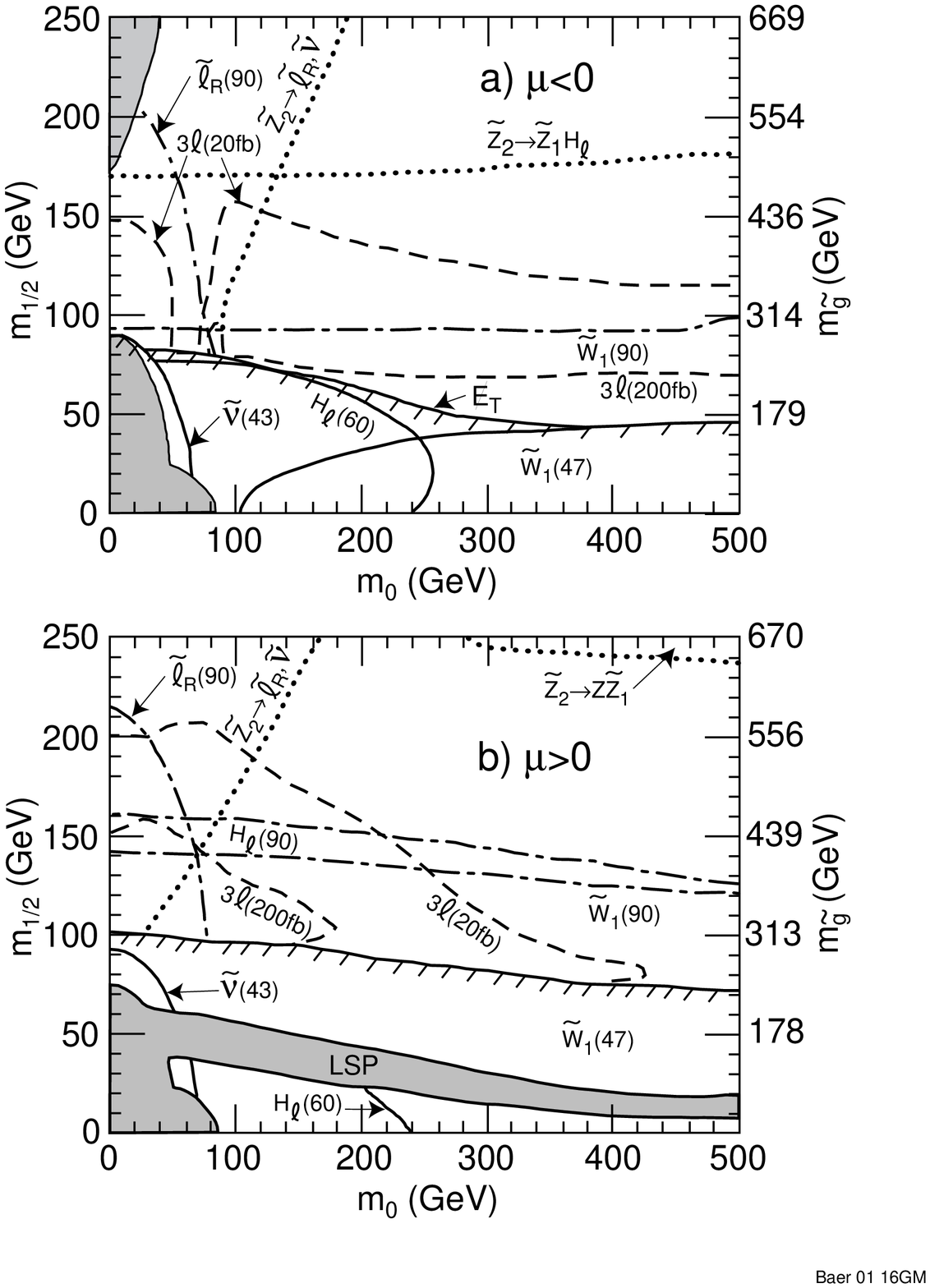,height=11cm}}
\caption[]{Regions in the $m_{0}\ vs. m_{1/2}$ plane explorable by
Tevatron and LEP II experiments.}
\label{fig64}
\end{figure}

Roughly then, LEP II is expected to probe slepton, chargino and squark masses
to $\sim 90$ GeV, as well as the light MSSM Higgs boson to $\sim 90$ GeV.
The $m_0\ vs.\ m_{1/2}$ plane seems to provide a
convenient panorama in which to plot results$^7$.
In Fig.~\ref{fig64}, we take $A_0=0$, $\tan\beta =2$ and $m_t=170$ GeV, and
show results
for both signs of $\mu$.
The gray regions are excluded on theoretical grounds, since in the lower
left region, electroweak symmetry breaking does not occur, or
cannot attain the proper $Z$-boson mass. Other parts of the
gray regions are excluded because
sparticles other than the $\tz_1$ become the LSP.
The current experimentally excluded region in Fig.~\ref{fig64}{\it a}
(shown by hatch marks) is due to four separate limits: the LEP limits that
$m_{\tw_1}>47$ GeV, $m_h \agt 60$ GeV, and $m_{\tnu}>43$ GeV, and
also the Tevatron $\etmiss +jets$ search.
In Fig.~\ref{fig64}{\it b}, the experimentally
excluded region is made up entirely of the LEP chargino mass bound.
Future searches at LEP II should probe below the dot-dashed lines,
which indicate the 90 GeV contours for $\tell$, $\tw_1$ and $H_{\ell}$.

\section{Sparticles at the Tevatron and its upgrades}

\subsection{$\tg\tg$, $\tg\tq$ and $\tq\tq$ signals}

Sparticle production at the Fermilab Tevatron collider is dominated by
$\tg\tg$, $\tg\tq$ and $\tq\tq$ production as long as $m_{\tg}<250-350$ GeV.
Once these sparticles are produced, they decay via cascades through
various charginos and neutralinos until the state containing the
lightest SUSY particle (LSP) is reached.
Thus, the final state consists of multiple leptons, jets and missing energy
from undetected LSP's and neutrinos. Searches at the CDF and D0
experiments have mainly focused on the multi-jet $+\eslt$ channel. By seeing
no signal above expected background rates, they conclude$^8$
(based on $\sim 10\ pb^{-1}$ of data) that $m_{\tg}\agt 150$ GeV if
$m_{\tq}\gg m_{\tg}$, or $m_{\tg}\agt 210$ GeV, if $m_{\tg}\sim m_{\tq}$.

The mass reach (in terms of $m_{\tg}$, for comparison with Fig. 1)
of the Tevatron experiments via various multi-isolated-lepton
topologies has been calculated in Ref. 9, for $100$ and $1000\ pb^{-1}$ of
integrated luminosity, for nominal cuts to remove the bulk of background,
and are listed in Table 1.
\begin{table}
\caption[]{Reach in $m_{\tg}$ via various event topologies
for the SUGRA-inspired MSSM, assuming an integrated
luminosity of 0.1 fb$^{-1}$ (1 fb$^{-1}$), at the Tevatron collider.
We use $m_t =150$ GeV for the background.}

\bigskip

\begin{tabular}{ccccccc}
case & $E\llap/_T$ & $1\ \ell$ & $OS$ & $SS$ & $3\ \ell$ & $\ge 4\ \ell$ \\
\hline
$m_{\tq}=m_{\tg}+10$ GeV & 240 (260) & --- (---) & 225 (290) &
230 (320) & 290 (425) & 190 (260) \\
$m_{\tq}=m_{\tg}-10$ GeV & 245 (265) & --- (---) & 160 (235) &
180 (325) & 240 (440) & --- (---) \\
$m_{\tq}=2m_{\tg}$ & 185 (200) & --- (---) & --- (180) &
160 (210) & 180 (260) & --- (---) \\
\end{tabular}
\end{table}

In the main injector (MI) era
(integrated luminosity $\sim 1\ fb^{-1}$), the reach in the $\etmiss +jets$
channel will be
background limited. However, in the $SS$ dilepton and $3\ell$ channels,
a much larger range of masses can be explored. In the $3\ell$ channel, for
$m_{\tq}\sim 2m_{\tg}$, $m_{\tg}\sim 260$ GeV can be explored, while for
$m_{\tq}\sim m_{\tg}$, squarks and gluinos as heavy as
425-440~GeV might be detectable.

\subsection{$\tw_1\tz_2\to 3\ell +\eslt$ signal}

Recently, there has been much interest in the $\tw_1\tz_2\to 3\ell +\eslt$
signal, due to its significant signal rate over much of parameter space,
and especially the low background rate expected from SM processes$^{10}$.
Using cuts designed to extract signal from background,
detailed simulations of this signal have recently been performed
over large regions
of SUGRA parameter space, for the Tevatron MI, and also a proposed luminosity
upgrade to $25\ fb^{-1}$ per year, dubbed $TeV^*$. Results are shown,
in a wonderful plot by Chen$^{10}$, for
$\mu <0$ and $\tan\beta =2$, in Fig. 2. The black squares indicate the reach
of Tevatron MI, while squares with X's and open squares designate
the $10\sigma$ and $5\sigma$ reach of $TeV^*$. In Fig. 2b, the corresponding
$\tg$, $\tw_1$ and $\tell_R$ mass contours are plotted. Thus, for the
low range of $m_0$, the MI can probe to $m_{\tg}\sim 450$ GeV, while
for large $m_0$, MI can probe to $m_{\tg}\sim 215-270$ GeV. The reach of
$TeV^*$ is considerably greater: $m_{\tg}\sim 700$ GeV for small $m_0$,
while $m_{\tg}\sim 500$ for large $m_0$. Unfortunately, this compelling
behavior doesn't persist if we switch the sign of $\mu$. For $\mu >0$,
there is an equivalent reach in terms of $m_{\tg}$ for small $m_0$, but for
large $m_0\agt 300-400$ GeV, interference effects in the
$\tz_2\to \ell\bar{\ell}\tz_1$ branching fraction kill the signal, giving
no reach at all via clean trileptons$^{10}$ for either MI, $TeV^*$, or
even LHC (see Fig. 1b $3\ell$ contour).
\begin{figure}
\centerline{\psfig{file=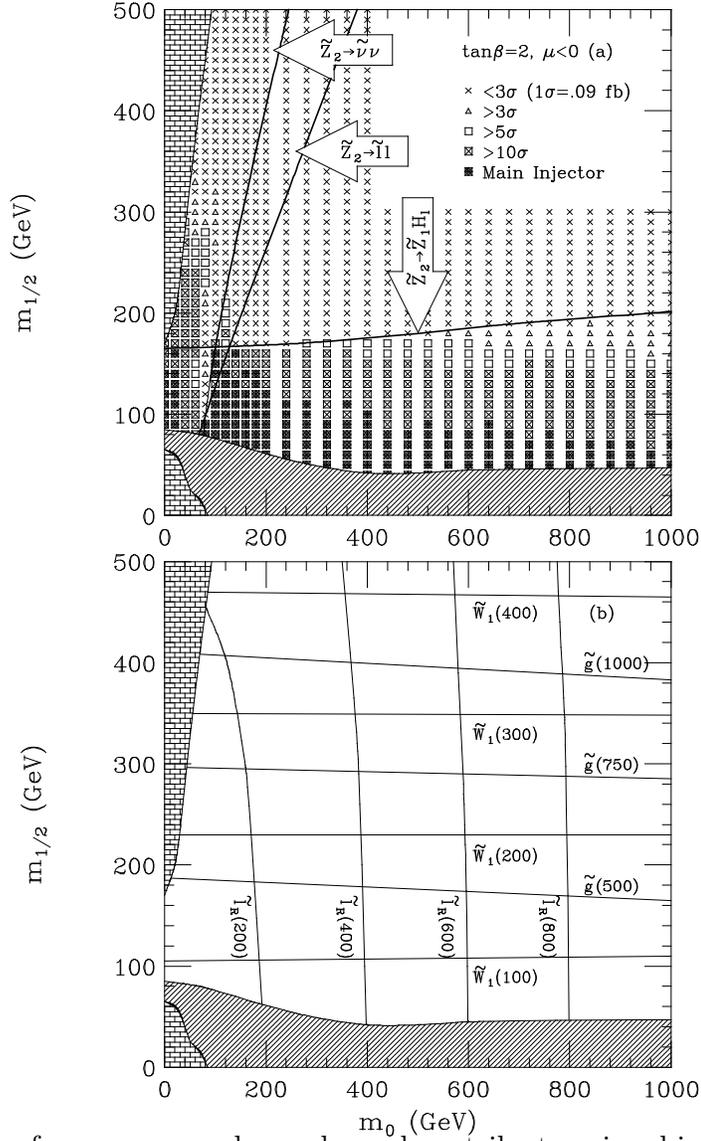,height=16cm,angle=90}}
\caption[]{Regions of $m_0\ vs.\ m_{1/2}$ plane where clean trilepton signal is
observable over background, for Tevatron MI project (1 $fb^{-1}$) and
TeV$^*$ (25 $fb^{-1}$). We take $A_0=0$ and $\mu <0$.}
\label{fig63}
\end{figure}

\subsection{Other sparticles}

The Tevatron also has the capability to search for light top squarks below
about $m_{\tst_1}\sim 100$ GeV$^{11}$.
If the decay $\tst\to b\tw_1$ is allowed, it should dominate top squark decays.
If so, top squark signatures are the same as top quark signatures, except
the decay products are softer. If the $\tst\to b\tw_1$ decay is
kinematically closed, then it is expected $\tst_1\to c\tz_1$ dominates.
In this case, one looks for charm dijet events, plus $\eslt$.

It appears to be difficult to search for sleptons at the Tevatron. The best
signature is in the dilepton $+\eslt$ channel. Unfortunately, a low signal rate
plus large backgrounds from $\tau\bar{\tau}$ and $WW$ production seem to
swamp this signal for $m_{\tell}\agt 50$ GeV$^{12}$.

\section{Sparticles at the LHC}

Sparticle production cross section at the LHC ($\sqrt{s}=14$ TeV) are
dominated by $\tg\tg$, $\tg\tq$ and $\tq\tq$ production.
Production of $\tg$ and $\tq$ is followed by decays through the usual
cascades. The multi-jets $+\eslt$ channel has been examined in detail
in the literature. (For more detailed discussion, see the talk by C. H. Chen,
these proceedings.) In particular, the Atlas collaboration finds the
reach capability for $m_{\tg}$ listed in Table 2$^{13}$.
\begin{table}
\caption[]{5$\sigma$
discovery limits in $m_{\tg}$ (in TeV) via $\etmiss +jets$ events
at LHC Atlas detector for various
choices of squark and gluino mass ratios, and collider integrated
luminosities. Variation of MSSM
parameters can cause these limits to vary by $\sim 150$ GeV.}

\bigskip
\centerline{
\begin{tabular}{cccc}
case & $10^3\ pb^{-1}$ & $10^4\ pb^{-1}$ & $10^5\ pb^{-1}$ \\
\hline
$m_{\tq}=m_{\tg}$ & 1.8 & 2.0 & 2.3 \\
$m_{\tq}=2m_{\tg}$ & 1.0 & 1.3 & 1.6 \\
\end{tabular}
}
\end{table}

These numbers have been confirmed by calculations in Ref. 14, where the mass
reach is also plotted in the $m_0\ vs.\ m_{1/2}$ plane. Furthermore, it
seems a rough measure of $m_{\tg}$ can be made to $15-25\% $, by reconstructing
hemispheric masses. Detailed calculations for other multi-lepton cascade
decay topologies are in progress.

What about other sparticles? It has been shown in Ref. 12 that LHC has a reach
in $m_{\tell}\sim 250$ GeV by looking for dilepton $+\eslt$ events.
Furthermore, LHC can see only slightly better than $TeV^*$ in the
$\tw_1\tz_2\to 3\ell +\eslt$ channel (in particular, the ``hole'' for
$\mu >0$ still persists at LHC). This is because the
$\tz_2\to\tz_1 H_{\ell}$ spoiler mode closes off a higher mass reach.
If such a signal is detected, then detailed information on
$m_{\tz_2}- m_{\tz_1}$ can be obtained.

\section{Conclusions}

In conclusion, only recently has the capability emerged to perform
realistic simulations of supersymmetry at hadron or $e^+e^-$ colliders.
We see that LEP II and Tevatron both will be able to explore significant
regions of parameter space for minimal SUGRA, and if they are lucky, may
discover it. However, of the approved facilities, only LHC can perform
a definitive search for sparticles below the TeV scale.
The capability of ultra-high energy linear $e^+e^-$ colliders was not discussed
here, but, depending on energy, may have comparable reach to LHC, and in
addition, be able to extract detailed and precise information on sparticle
masses and couplings$^{15}$.

\section{Acknowledgements}

I thank Chih Hao Chen, Jack Gunion, Chung Kao, Ray Munroe, Frank Paige,
Heath Pois, John Sender and Xerxes Tata for
collaborations on these various projects. In addition, I thank Jack Gunion
and all the UC-Davis workers for organizing a swell conference,
and for collaborations.

\end{document}